# sc-OTGM: Single-Cell Perturbation Modeling by Solving Optimal Mass Transport on the Manifold of Gaussian Mixtures

**Andac Demir, Elizaveta Solovyeva, James Boylan, Mei Xiao, Fabrizio Serluca, Sebastian Hoersch, Jeremy Jenkins, Murthy Devarakonda, Bulent Kiziltan**
Novartis, Biomedical Research

## Abstract

Influenced by recent breakthroughs in Large Language Models (LLMs), single-cell foundation models are emerging. While these models demonstrate successful performance in cell type clustering, phenotype classification, and gene perturbation response prediction, it remains to be seen if a simpler model could achieve comparable or better results, especially with limited data. This is important, as the quantity and quality of single-cell data typically fall short of the standards in textual data used for training LLMs. Single-cell sequencing often suffers from technical artifacts, dropout events, and batch effects. These challenges are compounded in a weakly supervised setting, where the labels of cell states can be noisy, further complicating the analysis. To tackle these challenges, we present sc-OTGM, streamlined with less than 500K parameters, making it approximately 100x more compact than the foundation models, offering an efficient alternative. sc-OTGM is an unsupervised model grounded in the inductive bias that the scRNA-seq data can be generated from a combination of the finite multivariate Gaussian distributions. The core function of sc-OTGM is to create a probabilistic latent space utilizing a Gaussian mixture model (GMM) as its prior distribution and distinguish between distinct cell populations by learning their respective marginal probability density functions (PDFs). It uses a Hit-and-Run Markov chain sampler to determine the optimal transport (OT) plan across these PDFs within the GMM framework. We evaluated our model against a CRISPR-mediated perturbation dataset, called CROP-seq, consisting of 57 one-gene perturbations. Our results demonstrate that sc-OTGM is effective in cell state classification, aids in the analysis of differential gene expression, and ranks genes for target identification through a recommender system. It also predicts the effects of single-gene perturbations on downstream gene regulation and generates synthetic scRNA-seq data conditioned on specific cell states. Source code and documentation are available at: https://github.com/Novartis/scOTGM.

## 1 Introduction

The molecular mechanisms that drive diseases are complex, often reflected in the high-dimensional profiles of gene expression. Conducting detailed analyses of these gene expression matrices—across various cell types, disease states, and control versus experimental subjects—is essential to understand disease progression and identify targets for potential drug interventions. scRNA-seq technology facilitates the detailed profiling of transcriptomes across a vast range of cells, from thousands to millions, allowing for the exploration of cellular heterogeneity and the understanding of disease pathogenesis at the single-cell level.

In recent years, Perturb-seq has emerged as a powerful high-throughput method that combines CRISPR-based genetic perturbation with scRNA-seq Dixit et al. (2016); Adamson et al. (2016). It enables the analysis of how specific gene modulations impact gene expression across numerous individual cells. Within this framework, CRISPRi and CRISPRa allows for targeted downregulation and upregulation of gene expression, respectively, offering insights into gene interactions and regulatory networks at the granularity of single-cells.

A significant hurdle in this research is that cellular responses to genetic perturbations are highly heterogeneous Elsasser (1984); Rubin (1990). This variability arises from differences in mRNA and protein levels Sonneveld et al. (2020), cell states Kramer et al. (2022), and the microenvironment among single-cells Snijder et al. (2009). Given the heterogeneity of potential perturbations, and the complexity of possible cell states, understanding the inherent data geometries and distributions of distinct cell populations becomes crucial for effective analyses. sc-OTGM employs a GMM to parametrize the marginal PDFs of diverse cell states and states within a reduced subspace. Not all subpopulations are well modeled by Gaussian distributions, and some subpopulations may cover large regions of feature space due to skewed class proportions and need further subdivisions. However, the use of GMM as priors enables the detection of local, high-density regions of phenotype space and effectively incorporates an inductive bias to overcome the limitations of data quantity and quality. This approach aligns with some recent methodologies in single-cell genomics, where Gaussian mixture priors are adopted to address the challenges of noise and data scarcity inherent in scRNA-seq datasets Wen et al. (2023); Xu et al. (2023); Grønbech et al. (2020); Slack et al. (2008).

While mixture modeling is not a new concept in scRNA-seq data analysis, we break new ground from following perspective: **First**, sc-OTGM effectively learns the OT plan that facilitates the mapping from one cell population to another on the manifold of Gaussian mixtures. To capture the transformation from unperturbed to perturbed cell states, the model employs a Hit-and-Run Markov chain sampler. This sampler guarantees a globally optimal solution to generate samples from the target distribution and offers faster convergence than small-step random walks Smith (1996). **Second**, modeling the PDF of perturbations allows us to identify perturbed genes and predict changes in the expression of other genes following perturbation. **Third**, sc-OTGM provides a scalable and unified framework for cell state classification, differential gene expression analysis, gene ranking for target identification, perturbation response prediction, and the generation of synthetic scRNA-seq data by sampling from the posteriors of Gaussian components. It offers improved efficiency with reduced runtime and memory requirements compared to LLMs and Variational Autoencoder (VAE) variants, while maintaining competitive accuracy.

## 2 Related Work

Recent advances in single-cell transcriptomics have led to the development of models such as Geneformer Theodoris et al. (2023), scGPT Cui et al. (2023), and scBERT Yang et al. (2022), which utilize masked language modeling (MLM) to learn gene embeddings. However, their effectiveness, particularly in zero-shot learning and in addressing batch effects, is still questioned Kedzierska et al. (2023); Boiarsky et al. (2023). Additionally, there have been significant developments in generative models, including Compositional Perturbation Autoencoder (CPA) Lotfollahi et al. (2021), which uses distinct encoder networks for cell states and perturbations, and further advancements by Lopez et al. (2023) and SAMS-VAE Bereket & Karaletsos (2023), which focus on disentangling representations with causal semantics for perturbation analysis. Tejada-Lapuerta et al. (2023) has criticized these approaches for oversimplifying complex causal relationships. Further details on these models and their comparative analyses can be found in the Appendix A.1.

## 3 Dataset

To evaluate the performance of sc-OTGM, we used the CROP-seq dataset from in-vitro experiments on human-induced pluripotent stem cell (iPSC)-derived neurons subjected to genetic perturbations Tian et al. (2021). These perturbations were executed via CRISPRi, enabling targeted gene knockdown to investigate its effects on neuronal survival and oxidative stress. Using the `rank_genes_groups` method from scanpy package Wolf et al. (2018) for differential expression analysis, we scrutinized the effects of knocking down 185 genes identified as potentially relevant to neuronal health and disease states. Of these, only 57 genes met our significance threshold (adjusted $p$-value less than 0.05), indicating a significant alteration in expression levels post-perturbation. These findings are summarized in Table 2, which includes the genes that presented significant differential expression, highlighting their potential roles in neuronal function and susceptibility to oxidative stress—a key factor in the pathogenesis of neurodegenerative diseases. Additional information regarding the pre-processing procedures can be found in Appendix A.4. Raw published data is available from the Gene Expression Omnibus under accession code GSE152988.

## 4 Generative Mixture Model

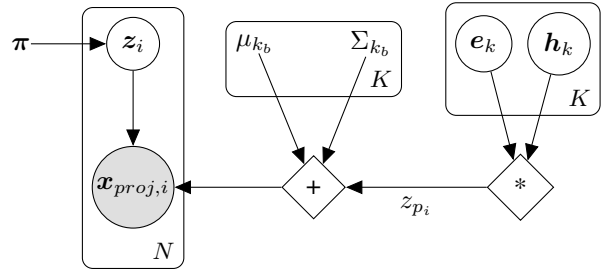

Figure 1: sc-OTGM represented as a generative graphical model.

We define the complete generative model for sc-OTGM as illustrated in Figure 1. Let $\mathbf{X}_{\text{proj}}$ denote the gene expression matrix, with rows representing individual cells and columns representing features in a reduced-dimensional space. The gene expression profile of cell $i$ is $\mathbf{X}_{\text{proj, i}}$. $\boldsymbol{\pi}$ represents the cluster probabilities. Each $\pi_k$, where $k$ specifies a particular cell state, indicates the prior probability of the $k$-th component in the mixture, subject to $\sum_k \pi_k = 1$. The latent variable $\boldsymbol{z}_i \in \mathbb{R}^K$ determining the component generating each data point $\boldsymbol{x}_{\text{proj, i}}$, is one-hot encoded and follows a categorical distribution parameterized by $\pi$. For each Gaussian component in the mixture model, $\mu_{k_b} \in \mathbb{R}^m$ and $\Sigma_{k_b} \in \mathbb{R}^{m \times m}$ define the mean and covariance matrix for unperturbed cells of a specific cell type, respectively. We specify perturbations and heterogeneous cellular responses as multivariate Gaussian-distributed variables $\boldsymbol{e} \sim \mathcal{N}(\mu_{k_e}, \Sigma_{k_e})$ and $\boldsymbol{h} \sim \mathcal{N}(\mu_{k_m}, \Sigma_{k_m})$, respectively. We model perturbation as a dynamic system, where the cell outputs an impulse response function $\boldsymbol{h}$, when presented with a brief perturbation signal $\boldsymbol{e}$. The convolution of these variables represents the combined effect on the latent state $\boldsymbol{z}_{p_i}$, which is also distributed as a multivariate Gaussian:

$$\boldsymbol{z}_{p_i} = (\boldsymbol{e}_k * \boldsymbol{h}_k)(z) = \int_{-\infty}^{+\infty} \boldsymbol{e}_k(\tau)\boldsymbol{h}_k(z-\tau)d\tau \sim \mathcal{N}(\mu_{k_e} + \mu_{k_m}, \Sigma_{k_e} + \Sigma_{k_m}) \sim \mathcal{N}(\mu_{k_p}, \Sigma_{k_p}), \quad (1)$$

In the proposed generative mixture model, the joint probability distribution for observed data $\mathbf{X}$, and latent variables $\boldsymbol{Z^i}$, $\boldsymbol{E}$, and $\boldsymbol{H}$ conditioned on $\boldsymbol{\pi}, \mu, \Sigma$ is formulated as:

$$p(\mathbf{X}, \boldsymbol{Z^i}, \boldsymbol{E}, \boldsymbol{H} \mid \boldsymbol{\pi}, \mu, \Sigma) = \left[ \prod_{i=1}^N p(\boldsymbol{z}_i | \boldsymbol{\pi}) p(\mathbf{x}_{proj,i} | \boldsymbol{z}_i, \mu, \Sigma) \right] \left[ \prod_{i=1}^N \boldsymbol{e} * \boldsymbol{h} \right] \quad (2)$$

$$= \prod_{i=1}^N \prod_{k=1}^K \left( \boldsymbol{\pi}_k^{\boldsymbol{z}_{ik}} \mathcal{N}(w_{b_i}; \mu_{k_b}, \Sigma_{k_b})^{\boldsymbol{z}_{ik}} \mathcal{N}(w_{p_i}; \mu_{k_p}, \Sigma_{k_p}) \right), \quad (3)$$

where $\boldsymbol{Z} \in \{0,1\}^{N \times K}$ denotes latent class indicators for $N$ cells across $K$ cell states, $\boldsymbol{E} \in \mathbb{R}^{N \times m}$ captures perturbation effects, and $\boldsymbol{H} \in \mathbb{R}^{N \times m}$ represents the cellular responses to these perturbations. Each datum $w_{b_i}, w_{p_i} \in \mathbb{R}^m$ is drawn independently and identically distributed (i.i.d.) from their respective marginal PDFs. To prevent numerical instability due to arithmetic underflow or overflow during likelihood calculations in the E-step of the Expectation-Maximization (EM) algorithm, we use the log-sum-exp trick. This method transforms the product of Gaussian probabilities into a sum, ensuring more stable computations. Let $L$ denote $p(\mathbf{X}, \boldsymbol{Z^i}, \boldsymbol{E}, \boldsymbol{H} \mid \boldsymbol{\pi}, \mu, \Sigma)$:

$$\log L = \sum_{i=1}^N \sum_{k=1}^K \left( \boldsymbol{z}_{ik} \log \boldsymbol{\pi}_k + \boldsymbol{z}_{ik} \log \mathcal{N}(w_{b_i}; \mu_{k_b}, \Sigma_{k_b}) + \log \mathcal{N}(w_{p_i}; \mu_{k_p}, \Sigma_{k_p}) \right) \quad (4)$$

The Maximum-a-Posteriori (MAP) parameter updates for the GMM via the EM algorithm are detailed in Algorithm 3. To address numerical instability in $\Sigma$'s inversion due to its near singularity or non-positive semi-definiteness, Tikhonov regularization is applied Alberti et al. (2021). See Appendix A.6 for additional details.

### 4.1 PLANNING OPTIMAL TRANSPORT VIA HIT-AND-RUN MARKOV CHAIN SAMPLER

OT problems, central to measuring the cost of optimally transporting mass from one distribution to another, are traditionally solved via the Monge (1781) and Kantorovich (1942) formulations, which, however, scale poorly for large datasets due to their reliance on linear programming (LP)Bunne et al. (2023). A breakthrough by Cuturi (2013) introduces entropic regularization into OT, resulting in the Sinkhorn algorithm, which significantly reduces computational complexity, enabling efficient large-scale applications. This methodological advancement, detailed in Section A.2.3, represents a pivotal shift towards practical OT computation in machine learning. *While OT is conventionally represented by a scalar value indicating the minimum cost required for such transport under specific constraints, sc-OTGM conceptualizes OT as a distribution to focus on the distribution of transportation costs and paths rather than summarizing these costs into a single scalar.* A distribution-based approach encapsulates more information about the transport process, such as the variance of transport costs, providing not just the minimum cost but also how costs are distributed across different transport paths. Additionally, it provides a more robust measure of similarity between distributions, as it does not collapse the transport problem into a single metric but rather considers the entire cost landscape, potentially mitigating the influence of outliers.

We model the latent states of unperturbed and perturbed cells as Gaussian distributions, with unperturbed cells described by $X = \mathcal{N}(\mu_{k_b}, \Sigma_{k_b})$ and perturbed cells by $Y = \mathcal{N}(\mu_{k_y}, \Sigma_{k_y})$. MAP estimates for $\mu_{k_b}, \Sigma_{k_b}, \mu_{k_y}, \Sigma_{k_y}$ were derived using the EM Algorithm within a GMM framework. To quantify the perturbation effect, we introduce a distribution, $Z = \mathcal{N}(\mu_{k_p}, \Sigma_{k_p})$, resulting from the linear displacement between $X$ and $Y$. Specifically, $Z$ is characterized by $\mu_{k_p} = \mu_{k_y} - \mu_{k_b}$ and $\Sigma_{k_p} = \Sigma_{k_b} + \Sigma_{k_y} - 2\Sigma_{\text{cross}}$. $Z$ captures the OT cost distribution required to transition between these states. The coupling (joint distribution) of $X$ and $Y$ is unknown, therefore we approximate $\Sigma_{\text{cross}}$ via Hit-and-Run Markov Chain Monte Carlo (MCMC). This generates a Markov chain that, in its stable state, converges to the uniform distribution over a convex polytope van Valkenhoef & Tervonen (2015), and under certain regularity conditions, converges in distribution to the target distribution Smith (1996). The steps to compute $\Sigma_{\text{cross}}$ are elaborated in Algorithm 1. For recursive updates of $\Sigma_{\text{cross}}$ we use follow-the-leader (FTL) strategy which is prominently used in online density estimation and active learning Azoury & Warmuth (2001); Dasgupta & Hsu (2007). Details on the implementation and synthetic data experiments are provided in the Appendix A.8.

sc-OTGM samples transportation paths (vectors) directly from the OT landscape, rooted in the perturbation distribution $Z$, within a dimensionally reduced subspace: $x_{\text{path}} \sim \mathcal{N}(\mu_{k_p}, \Sigma_{k_p})$, which effectively captures the essence of gene expression dynamics under perturbation. The high-dimensional gene expression profiles corresponding to these paths are reconstructed via inverse PCA, expressed as: $x''_{\text{path}} = x_{\text{path}} V_m^T$, where $V_m$ represents the matrix comprising the top $m$ eigenvectors derived from the covariance matrix of the pre-processed gene expression data. See Appendix A.4 and A.5 for more details. For each gene expression profile within the OT landscape, we derive a parametric representation, as follows:

$$\mu_{\text{gene},i} = \text{the } i^{\text{th}} \text{ element of } (\mu_{k_p} V_m^T) \quad \text{and} \quad \sigma^2_{\text{gene},i} = \text{the } i^{\text{th}} \text{ diagonal entry of } (V_m \Sigma_{k_p} V_m^T),$$

where $\mu_{\text{gene},i}$ denotes the degree of change in a gene's expression in response to perturbation, indicating potential activation or suppression of the gene. $\sigma_{\text{gene},i}$ quantifies the confidence in these expression changes, providing insights into the variability of our computations. While biological systems are often complex and nonlinear, under specific conditions or within certain ranges, linear approximations can provide valuable insights and simplify modeling efforts. van Someren et al. (2000) presents a methodology for modeling genetic networks that employs clustering to tackle the dimensionality problem and a linear model to infer the regulatory interactions. By analyzing the covariance matrix $\Sigma_{k_p}$ of the perturbation distribution, we can explore the interconnected behavior between genes, examining how they co-vary or influence each other. Consider the scenario where the expression level of gene $i$ changes, denoted as $\Delta X_i$. The covariance matrix $\Sigma_{k_p}$, upon transformation via $V_m$, becomes $\Sigma''_{k_p} = V_m \Sigma_{k_p} V_m^T$. The expected change in gene $j$'s expression level, $\Delta X_j$, resulting from a change in gene $i$, is linearly approximated as follows:

$$\Delta X_j = \frac{\Sigma''_{k_p, ij}}{\Sigma''_{k_p, ii}} \Delta X_i \tag{5}$$

---

**Algorithm 1** Estimation of Cross-Covariance Matrix via Hit-and-Run Markov Chain Monte Carlo

---

1: **Input:** Domains $\mathcal{D}_X$ and $\mathcal{D}_Y$ with non-zero density for $X$ and $Y$, number of iterations $N$, mean vectors $\mu_X$, $\mu_Y$, covariance matrices $\Sigma_X$, $\Sigma_Y$, and confidence interval for the bounds $\alpha$.
2: **Output:** Estimated cross-covariance matrix $\Sigma_{XY}$.
3: Initialize $(x^{(0)}, y^{(0)})$ uniformly from $\mathcal{D}_X \times \mathcal{D}_Y$.
4: Initialize $\Sigma_{XY}^{(0)}$ as a random matrix, ensuring it is symmetric and positive definite.
5: Compute the z-score, $z$, for the specified confidence interval $\alpha$ using the inverse of the standard normal CDF: $z = \Phi^{-1}\left(\frac{1+\alpha}{2}\right)$.
6: **for** $i = 1$ to $N$ **do**
7:     **Calculate non-zero density bounds $[a_X, b_X]$ for $X$:**
8:         Generate a random unit direction $\mathbf{d}_X$ in $X$'s space.
9:         Normalize $\mathbf{d}_X$ to unit length: $\mathbf{d}_{X,\text{normalized}} = \frac{\mathbf{d}_X}{\|\mathbf{d}_X\|}$
10:        Project $x^{(i-1)}$ onto $\mathbf{d}_{X,\text{normalized}}$: $p_{X,\text{projection}} = x^{(i-1)} \cdot \mathbf{d}_{X,\text{normalized}}$
11:        Compute standard deviation $\sigma_{X,\text{projection}} = \sqrt{\mathbf{d}_{X,\text{normalized}}^T \Sigma_X \mathbf{d}_{X,\text{normalized}}}$
12:        Determine $[a_X, b_X]$ using $p_{X,\text{projection}} \pm z \cdot \sigma_{X,\text{projection}}$
13:     **Calculate non-zero density bounds $[a_Y, b_Y]$ for $Y$:**
14:         Generate a random unit direction $\mathbf{d}_Y$ in $Y$'s space.
15:        Normalize $\mathbf{d}_Y$ to unit length: $\mathbf{d}_{Y,\text{normalized}} = \frac{\mathbf{d}_Y}{\|\mathbf{d}_Y\|}$
16:        Project $y^{(i-1)}$ onto $\mathbf{d}_{Y,\text{normalized}}$: $p_{Y,\text{projection}} = y^{(i-1)} \cdot \mathbf{d}_{Y,\text{normalized}}$
17:        Compute standard deviation $\sigma_{Y,\text{projection}} = \sqrt{\mathbf{d}_{Y,\text{normalized}}^T \Sigma_Y \mathbf{d}_{Y,\text{normalized}}}$
18:        Determine $[a_Y, b_Y]$ using $p_{Y,\text{projection}} \pm z \cdot \sigma_{Y,\text{projection}}$
19:     **Update $\Sigma_{XY}$:**
20:        Sample $x^{(i)} \sim \text{Uniform}(a_X, b_X)$   and   Sample $y^{(i)} \sim \text{Uniform}(a_Y, b_Y)$
21:        $\Delta x^{(i)} = x^{(i)} - \mu_X$, $\Delta y^{(i)} = y^{(i)} - \mu_Y$
22:        $\Sigma_{XY}^{(i)} = \frac{i}{i+1} \Sigma_{XY}^{(i-1)} + \frac{i}{(i+1)^2} \Delta x^{(i)} (\Delta y^{(i)})^T$       ▷ FTL Online Density Estimation
23: **end for**

---

# 5 Results

Not all cells receiving the CRISPR components will have the target gene successfully knockeddown, and even among those that do, the degree of knockdown can vary substantially due to differences in Cas9 activity, guide RNA efficiency, and individual cell responses. Additionally, the downstream effects of knocking down a particular gene leads to compensatory mechanisms in the cell, altering the expression of other genes. This necessitates benchmarking computational models based on their accuracy to identify knockeddown genes and predict subsequent changes in the expression of other genes. We evaluated statistical models for differential gene expression in CRISPRi experiments, including the Mann-Whitney U test, t-test, and sc-OTGM, as shown in Table 1. We assessed how often each method placed the true knockeddown gene within the top-k results. Lower p-values from the Mann-Whitney U test and t-test correlate with higher rankings. sc-OTGM performs exceptionally well in Top-1 accuracy, showing its effectiveness in identifying the most likely perturbed gene. The performance advantage of sc-OTGM decreases as the ranking threshold increases.

Table 1: Benchmarking differential gene expression analysis techniques. The highest performance is highlighted in **bold**, the best baseline method is <u>underlined</u>.

| Diff. Expression Analysis | Top-1 Acc. ↑ | Top-5 Acc. ↑ | Top-10 Acc. ↑ | Top-50 Acc. ↑ | Top-100 Acc. ↑ |
|---|---|---|---|---|---|
| Mann–Whitney U test | 0.37 | 0.40 | 0.42 | 0.58 | 0.60 |
| t-test | <u>0.39</u> | <u>0.67</u> | <u>0.70</u> | <u>0.79</u> | <u>0.86</u> |
| sc-OTGM | **0.56** | **0.68** | **0.74** | **0.82** | **0.91** |

In our experiments, we investigated the impact of 57 single-gene knockdowns, featuring BIN1 as a case study in the paper. The dataset for this case study, comprising 382 cells (315 control and 67 target), was split into an 80-20 train-test ratio. The extremely limited size of the dataset presented significant challenges in training our model from scratch and fine-tuning existing foundation models for the CRISPRi experiment. Figure 2 shows both sc-OTGM's perturbation distribution and a confusion matrix for cell state classification. Here, clustering around the center suggests minimal impact on most genes, whereas known differentially expressed genes (DEGs) (blue points) significantly diverge. sc-OTGM accurately ranked BIN1 as a knockeddown gene, placing it at the top of the recommendation list. In the presented confusion matrix, cells are classified as either Control—indicating no CRISPR-induced changes—or Perturbed—where changes are expected. True negatives (56) correctly identify cells as Control; false positives (18) erroneously label cells as Perturbed; false negatives (5) miss the classification of Perturbed cells as Control; and true positives (8) correctly detect Perturbed cells. Additionally, we used in silico perturbation response predictions to analyze gene expression changes between BIN1-knockdown and control cells, as shown in Figure 3. Our results confirm that sc-OTGM accurately predicts the direction of expression level changes among DEGs, distinguishing between upregulated and downregulated genes following BIN1 knockdown. However, the differences between the magnitudes of model predictions and actual values highlight the potential advantages of nonlinear models in capturing the complex dynamics of gene regulation. Furthermore, based on sc-OTGM's ranking, a cutoff of 100 is set to select the top predicted DEGs. Predicted DEGs are obtained from the top of the ranked gene list, and compared against the list of known DEGs. We conducted Fisher's exact test which yielded a $p$-Value of $6.39e-08$, significantly below the standard threshold of 0.05. This confirms a strong statistical correlation between known DEGs after BIN1 knockdown and those predicted by sc-OTGM. While this analysis focuses on BIN1, Table 3 presents results for all targeted gene knockdowns, showing $p$-Values well below 0.05, validating the performance on DEG enrichment. Additionally, sc-OTGM demonstrated a mean accuracy of $75.2\% \pm 15.4\%$ and F1-score of $0.79 \pm 0.14$, predicting the directionality of expression changes (upregulation or downregulation) in DEGs.

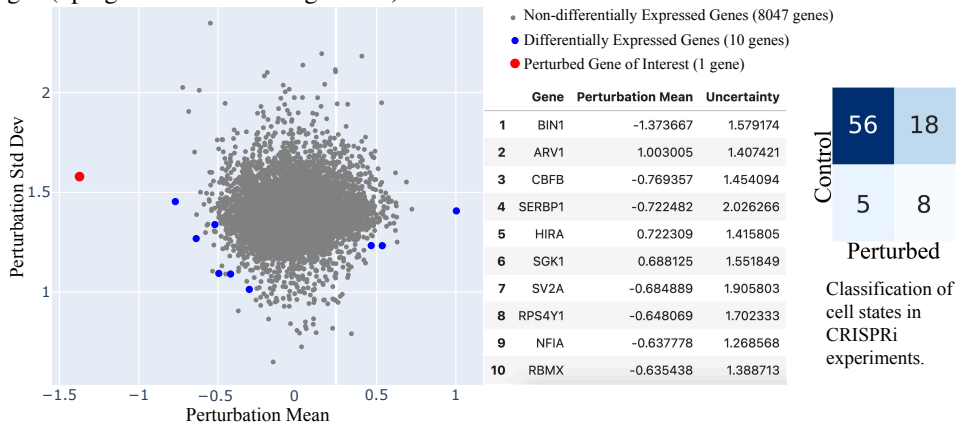

Figure 2: Perturbation distribution from the OT landscape, and ranking genes for target identification through a recommender system.

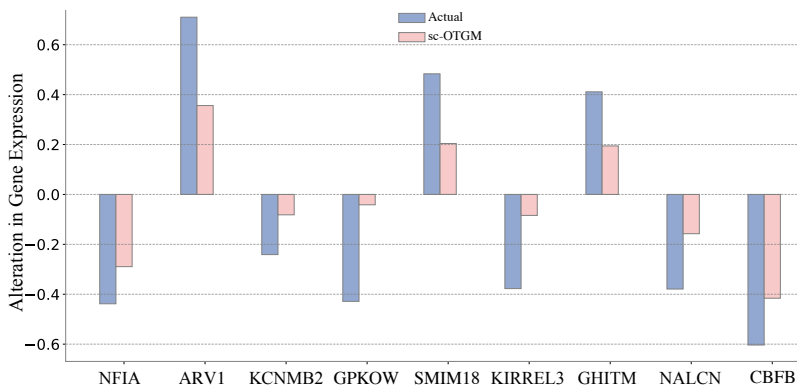

Figure 3: In silico perturbation response prediction of DEGs following the knockdown of BIN1.

# 6 CONCLUSION

sc-OTGM framework efficiently detects genes with significant activity changes, providing an alternative to single-cell foundation models in data-limited, high-noise scenarios, alleviating the need for extensive fine-tuning. Additionally, it can identify up/down-regulated genes post-perturbation without establishing causality.

# A APPENDIX

## A.1 RELATED WORK

Influenced by recent breakthroughs in LLMs, single-cell foundation models like scBERT Yang et al. (2022), scGPT Cui et al. (2023), and Geneformer Theodoris et al. (2023) have emerged as potential tools in single-cell biology, demonstrating successful performance in cell type clustering, phenotype classification, and gene perturbation response prediction. scBERT randomly masks a fraction of non-zero gene expression values and predicts them based on the remaining data. scGPT introduces a variant of masked language modeling (MLM) that mimics the auto-regressive generation in natural language processing, where the masked genes are iteratively predicted according to the model's confidence. Geneformer completely abandons the precise expression levels of genes. Instead, it models the rank of gene expressions and constructs sequences of genes according to their relative expression levels within each cell. The assessment of Geneformer and scGPT revealed significant limitations in their zero-shot performance Kedzierska et al. (2023). In various tasks, particularly in cell type annotation, these models are outperformed by simpler models, such as scVI Lopez et al. (2018) and strategies focusing on highly variable genes (HVGs). Kedzierska et al. (2023) also highlighted that aligning the pretraining dataset's tissue of origin with the target task did not consistently improve scGPT's performance, and broader pretraining datasets sometimes resulted in decreased effectiveness. Additionally, both Geneformer and scGPT showed inadequate handling of batch effects in zero-shot settings. Comparisons of scBERT and scGPT with L1-regularized logistic regression in cell type annotation under limited training data suggest that logistic regression performs more accurately, questioning the complexity needed for cell type annotation and the efficacy of MLM in learning gene embeddings Boiarsky et al. (2023). The failure of these models to accurately predict gene expression in zero-shot or limited data scenarios emphasizes the necessity for advancements in model design and training methodologies. CellPLM aggregates gene embeddings since gene expressions are bag-of-word features and leverages spatially-resolved transcriptomic (SRT) data in pre-training to facilitate learning cell-cell relations and introduce a Gaussian mixture prior as an additional inductive bias to overcome data limitation Wen et al. (2023).

The field has also seen advancements in generative modeling. Compositional Perturbation Autoencoder (CPA) has been developed to learn embeddings for both cell states and perturbations Lotfollahi et al. (2021). This is achieved within a unified framework, utilizing input data comprising two main components: gene expression data and perturbation data, such as drug types and dosages. CPA employs distinct encoder networks for cell states and perturbations, respectively. These encoders map input data into a latent space, from which the decoder network reconstructs the expected cellular response to a specific perturbation. Building on that concept, Lopez et al. (2023) leverages sparse mechanism shift assumption in order to learn disentangled representations with a causal semantic to the analysis of single-cell genomics data with genetic or chemical perturbations. However, in single-cell transcriptomics, such a setting might be overly simplistic. Similarly, SAMS-VAE adopts the idea to disentangle cellular latent spaces into basal and perturbation latent variables Bereket & Karaletsos (2023). Specifically, the proposed method models the latent state of perturbed samples as a combination of a local latent variable capturing sample-specific variation and sparse global variables of latent intervention. These global variables are formulated as the point-wise product between latent perturbation variables and a binary mask. Tejada-Lapuerta et al. (2023) argues that disentanglement simplifies the problem of recovering meaningful causal representations assuming independence among latent variables by the use of mean field approximation to ease the computation of variational inference.

## A.2 OPTIMAL TRANSPORT

### A.2.1 MONGE'S FORMULATION

Monge's formulation of the OT problem seeks a mapping $T : X \rightarrow Y$ that transports a mass distribution $\mu$ on a space $X$ to a mass distribution $\nu$ on a space $Y$ in the most cost-effective way. The cost of transporting a unit mass from point $x \in X$ to point $y \in Y$ is given by a cost function $c(x, y)$. The goal is to minimize the total transportation cost:

$$\min_T \int_X c(x, T(x)) \, d\mu(x)$$

subject to the constraint that $T_\# \mu = \nu$, where $T_\# \mu$ is the pushforward measure of $\mu$ by $T$, ensuring that the mass distribution after transportation is $\nu$.

### A.2.2 Kantorovich's Relaxation

Kantorovich's formulation relaxes Monge's problem by considering a probabilistic coupling $\pi$ in the product space $X \times Y$, which represents a joint distribution of source and target points that respects the marginal distributions $\mu$ and $\nu$. The problem is formulated as:

$$\min_{\pi \in \Pi(\mu,\nu)} \int_{X \times Y} c(x,y) \, d\pi(x,y)$$

where $\Pi(\mu, \nu)$ is the set of all couplings $\pi$ with marginals $\mu$ on $X$ and $\nu$ on $Y$. This formulation is more flexible than Monge's because it allows for mass splitting, making it possible to find solutions in situations where Monge's problem has none.

The Kantorovich problem leads to a dual formulation that expresses the OT cost as:

$$\sup_{(f,g) \in \Phi} \left\{ \int_X f(x) \, d\mu(x) + \int_Y g(y) \, d\nu(y) \right\}$$

where $\Phi$ consists of all pairs of functions $(f, g)$ such that $f(x) + g(y) \leq c(x, y)$ for all $(x, y) \in X \times Y$.

The $p$-Wasserstein distance between $\mu$ and $\nu$ for $p \geq 1$ is derived from Kantorovich's problem with the cost function $c(x, y) = \|x - y\|^p$, providing a metric on the space of probability measures:

$$W_p(\mu, \nu) = \left( \min_{\pi \in \Pi(\mu,\nu)} \int_{X \times Y} \|x - y\|^p \, d\pi(x,y) \right)^{1/p}$$

This distance measures the minimal amount of work required to transform the distribution $\mu$ into the distribution $\nu$ under the given cost function.

### A.2.3 Regularized Transport with Sinkhorn's Algorithm

The Sinkhorn distance introduces an entropic regularization to the OT problem, making it computationally more tractable by allowing the use of efficient algorithms. The entropic regularization term, $\epsilon \, \mathrm{KL}(\pi \| \mu \otimes \nu)$, where $\epsilon$ is a positive regularization parameter and KL denotes the Kullback-Leibler divergence between $\pi$ and the product distribution $\mu \otimes \nu$, adds strong convexity to the optimization problem. This convexity ensures that the smoothness of the objective function allows for the application of gradient-based optimization methods that are known to converge quickly.

$$\min_{\pi \in \Pi(\mu,\nu)} \int_{X \times Y} c(x,y) \, d\pi(x,y) + \epsilon \, \mathrm{KL}(\pi \| \mu \otimes \nu)$$

### A.3 Dataset Statistics

Table 2: Quantitative Analysis of Gene Expression Alterations Post-CRISPRi Knockdown

| Gene | $\log_2$(fold change) | $p$-Value | Adjusted $p$-Value | Num. Control Cells | Num. Targeted Cells |
|------|------|------|------|------|------|
| TUBB4A | $-5.60$ | $1.30 \times 10^{-29}$ | $1.06 \times 10^{-25}$ | 317 | 86 |
| ATP1A3 | $-5.57$ | $1.11 \times 10^{-25}$ | $9.08 \times 10^{-22}$ | 354 | 87 |
| KIFAP3 | $-5.34$ | $1.50 \times 10^{-26}$ | $1.23 \times 10^{-22}$ | 358 | 100 |
| MAPT | $-5.13$ | $2.26 \times 10^{-23}$ | $1.84 \times 10^{-19}$ | 368 | 101 |
| CASP3 | $-4.73$ | $1.53 \times 10^{-20}$ | $1.25 \times 10^{-16}$ | 307 | 83 |
| APEX1 | $-4.72$ | $8.08 \times 10^{-18}$ | $6.60 \times 10^{-14}$ | 346 | 76 |
| COX10 | $-4.56$ | $5.61 \times 10^{-14}$ | $3.28 \times 10^{-10}$ | 97 | 55 |
| NDUFS8 | $-4.52$ | $1.57 \times 10^{-15}$ | $1.29 \times 10^{-11}$ | 352 | 73 |
| ZNF292 | $-4.36$ | $4.91 \times 10^{-13}$ | $4.01 \times 10^{-9}$ | 339 | 60 |

Table 2 – *Continued from previous page*

| Gene | $\log_2$(fold change) | $p$-Value | Adjusted $p$-Value | Num. Control Cells | Num. Targeted Cells |
|---|---|---|---|---|---|
| GSTA4 | $-4.12$ | $3.74 \times 10^{-14}$ | $3.06 \times 10^{-10}$ | 337 | 73 |
| STX1B | $-4.01$ | $3.34 \times 10^{-16}$ | $2.72 \times 10^{-12}$ | 235 | 72 |
| OPTN | $-3.95$ | $4.06 \times 10^{-16}$ | $3.32 \times 10^{-12}$ | 338 | 88 |
| SOD1 | $-3.84$ | $3.71 \times 10^{-15}$ | $3.03 \times 10^{-11}$ | 364 | 85 |
| NDUFV1 | $-3.72$ | $8.89 \times 10^{-11}$ | $5.80 \times 10^{-7}$ | 320 | 62 |
| CALB1 | $-3.62$ | $4.02 \times 10^{-7}$ | $8.21 \times 10^{-4}$ | 50 | 64 |
| EEF2 | $-3.60$ | $2.61 \times 10^{-12}$ | $5.32 \times 10^{-9}$ | 367 | 63 |
| BIN1 | $-3.57$ | $3.12 \times 10^{-11}$ | $2.55 \times 10^{-7}$ | 315 | 67 |
| SCFD1 | $-3.56$ | $2.38 \times 10^{-6}$ | $3.60 \times 10^{-4}$ | 249 | 32 |
| PON2 | $-3.50$ | $6.35 \times 10^{-11}$ | $5.18 \times 10^{-7}$ | 99 | 74 |
| BAX | $-3.45$ | $3.23 \times 10^{-14}$ | $2.63 \times 10^{-10}$ | 230 | 86 |
| SCAPER | $-3.07$ | $1.37 \times 10^{-9}$ | $1.12 \times 10^{-5}$ | 311 | 71 |
| CYB561 | $-3.06$ | $1.15 \times 10^{-5}$ | $1.57 \times 10^{-3}$ | 138 | 32 |
| AKAP9 | $-3.00$ | $1.13 \times 10^{-9}$ | $9.21 \times 10^{-6}$ | 366 | 79 |
| VPS35 | $-2.97$ | $1.38 \times 10^{-8}$ | $3.76 \times 10^{-5}$ | 334 | 65 |
| PRNP | $-2.92$ | $8.65 \times 10^{-9}$ | $2.35 \times 10^{-5}$ | 253 | 72 |
| AP2A2 | $-2.87$ | $2.51 \times 10^{-11}$ | $2.05 \times 10^{-7}$ | 332 | 99 |
| SOD2 | $-2.80$ | $6.16 \times 10^{-8}$ | $5.03 \times 10^{-5}$ | 188 | 61 |
| BECN1 | $-2.79$ | $7.42 \times 10^{-6}$ | $1.89 \times 10^{-2}$ | 148 | 27 |
| SNCB | $-2.71$ | $3.35 \times 10^{-8}$ | $9.12 \times 10^{-5}$ | 144 | 68 |
| CDH11 | $-2.71$ | $1.55 \times 10^{-6}$ | $1.58 \times 10^{-3}$ | 97 | 61 |
| ELOVL5 | $-2.66$ | $1.36 \times 10^{-9}$ | $1.11 \times 10^{-5}$ | 173 | 92 |
| NTRK2 | $-2.66$ | $2.21 \times 10^{-4}$ | $1.62 \times 10^{-2}$ | 237 | 31 |
| DAP | $-2.65$ | $8.90 \times 10^{-8}$ | $3.63 \times 10^{-4}$ | 156 | 70 |
| EIF4G1 | $-2.49$ | $1.17 \times 10^{-6}$ | $1.63 \times 10^{-3}$ | 281 | 68 |
| TRPM7 | $-2.46$ | $5.46 \times 10^{-7}$ | $4.46 \times 10^{-3}$ | 100 | 85 |
| COASY | $-2.37$ | $1.06 \times 10^{-7}$ | $4.32 \times 10^{-4}$ | 120 | 101 |
| TRAP1 | $-2.34$ | $1.82 \times 10^{-7}$ | $7.44 \times 10^{-4}$ | 206 | 86 |
| CYP46A1 | $-2.32$ | $5.03 \times 10^{-7}$ | $4.11 \times 10^{-3}$ | 163 | 86 |
| PARP1 | $-2.25$ | $2.66 \times 10^{-7}$ | $9.08 \times 10^{-4}$ | 320 | 88 |
| FOXRED1 | $-2.25$ | $4.29 \times 10^{-5}$ | $1.78 \times 10^{-2}$ | 139 | 53 |
| AFG3L2 | $-2.24$ | $2.13 \times 10^{-6}$ | $8.69 \times 10^{-3}$ | 274 | 79 |
| RAB7A | $-2.14$ | $9.64 \times 10^{-8}$ | $3.94 \times 10^{-4}$ | 344 | 95 |
| PPP2R2B | $-2.13$ | $8.95 \times 10^{-6}$ | $8.84 \times 10^{-3}$ | 324 | 77 |
| RGS2 | $-2.10$ | $1.73 \times 10^{-5}$ | $1.09 \times 10^{-2}$ | 118 | 78 |
| AMFR | $-2.08$ | $1.61 \times 10^{-4}$ | $4.88 \times 10^{-2}$ | 96 | 69 |
| MRPL10 | $-2.06$ | $7.03 \times 10^{-5}$ | $1.34 \times 10^{-2}$ | 127 | 69 |
| ANO10 | $-1.94$ | $1.86 \times 10^{-5}$ | $8.91 \times 10^{-3}$ | 145 | 92 |
| DMXL1 | $-1.94$ | $6.13 \times 10^{-5}$ | $2.15 \times 10^{-2}$ | 107 | 89 |
| HYOU1 | $-1.91$ | $2.91 \times 10^{-5}$ | $1.83 \times 10^{-2}$ | 177 | 86 |
| HTT | $-1.89$ | $9.44 \times 10^{-7}$ | $3.85 \times 10^{-3}$ | 200 | 125 |
| ECHS1 | $-1.88$ | $2.57 \times 10^{-6}$ | $6.12 \times 10^{-3}$ | 332 | 96 |
| CYCS | $-1.85$ | $1.45 \times 10^{-6}$ | $3.04 \times 10^{-3}$ | 365 | 65 |
| CEP63 | $-1.84$ | $3.53 \times 10^{-5}$ | $1.92 \times 10^{-2}$ | 158 | 87 |
| FARP1 | $-1.80$ | $6.95 \times 10^{-5}$ | $3.14 \times 10^{-2}$ | 327 | 74 |
| FRMD4A | $-1.72$ | $7.80 \times 10^{-4}$ | $2.36 \times 10^{-2}$ | 246 | 67 |
| RPL6 | $-1.66$ | $3.29 \times 10^{-9}$ | $2.69 \times 10^{-5}$ | 368 | 87 |
| PFN1 | $-1.50$ | $2.87 \times 10^{-5}$ | $2.93 \times 10^{-2}$ | 366 | 76 |

## A.4  PRE-PROCESSING

**Cell Filtering** Cells expressing fewer than $\theta_g$ genes are excluded. Formally, cell $i$ is removed if its gene count $\sum_j \mathbf{1}_{\{X_{ij} > 0\}}$ is less than $\theta_g$, where $X_{ij}$ denotes the expression level of gene $j$ in cell $i$, and $\mathbf{1}$ represents the indicator function.

**Gene Filtering** Similarly, genes expressed in fewer than $\theta_c$ cells are discarded. A gene $j$ is eliminated if $\sum_i \mathbf{1}_{\{X_{ij}>0\}}$ falls below $\theta_c$.

**Normalization and Logarithmic Scaling** The gene expression dataset undergoes normalization to equalize expression levels across cells, parameterized by `counts_per_cell_after`. Post-normalization, logarithmic scaling is applied: $X'_{ij} = \log(1 + X_{ij})$, where $X_{ij}$ represents the normalized expression of gene $j$ in cell $i$.

**Identification of Highly Variable Genes** Genes are deemed highly variable based on their mean expression $\mu_j$ and dispersion $\sigma_j^2$, constrained within specified thresholds. A gene $j$ qualifies as highly variable if it meets the criteria $\theta_{\mu,\min} < \mu_j < \theta_{\mu,\max}$ and $\sigma_j^2 > \theta_\sigma$.

**Scaling** Subsequently, gene expression levels are standardized, ensuring zero mean and unit variance for each gene: $X'_{ij} = \frac{X_{ij} - \overline{X}_j}{\sigma_j}$, where $\overline{X}_j$ and $\sigma_j$ denote the mean and standard deviation of gene $j$'s expression, respectively.

**Final Steps** After the selection of highly variable genes, further scaling is applied: $X''_{ij} = \min(\max(X'_{ij}, -\theta_{\max}), \theta_{\max})$. This step ensures that the data range remains within the predefined limits, specified by $\theta_{\max}$.

## A.5 Projection to Reduced Subspace

In downstream analysis, Principal Component Analysis (PCA) is utilized to reduce the dimensionality of a gene expression matrix, $\mathbf{X}'' \in \mathbb{R}^{N \times d}$, comprising $N$ cells and $d$ genes. This process involves the eigendecomposition of the covariance matrix $\Sigma$ of pre-processed data $\mathbf{X}''$, to compute eigenvectors $v_i$ and their corresponding eigenvalues $\lambda_i$, satisfying the equation $\Sigma v_i = \lambda_i v_i$. The top $m$ eigenvectors, selected based on the magnitude of their eigenvalues, are represented as a matrix $V_m \in \mathbb{R}^{d \times m}$. The data $\mathbf{X}''$ is then projected onto the lower-dimensional subspace defined by $V_m$, resulting in the projected data $\mathbf{X}_{\text{proj}} = \mathbf{X}'' V_m$, where $\mathbf{X}_{\text{proj}} \in \mathbb{R}^{N \times m}$ and $m \ll d$.

---

**Algorithm 2** K-S Test for Each Principal Component of a High-Dimensional Sample against a Gaussian Reference Distribution

---

1: **Null Hypothesis** $H_0$: The data for principal component $i$ follows the reference distribution $f(x)$.
2: **Significance Level** $\alpha$: Threshold ($\alpha = 0.05$) to decide on the null hypothesis based on the K-S statistic.
3: **for** each principal component $i$ in $\boldsymbol{x}_{\text{proj}} \in \mathbb{R}^{N \times m}$ **do**
4:     Extract all data points in dimension $i$ to form $S_i = \{x_{\text{proj}}^{1,i}, x_{\text{proj}}^{2,i}, \dots, x_{\text{proj}}^{N,i}\}$
5:     Compute mean $\mu_i$ and variance $\sigma_i^2$ of $S_i$
6:     Define the Gaussian reference distribution $f(x) = \mathcal{N}(x; \mu_i, \sigma_i^2)$
7:     Sort $S_i$ in ascending order
8:     **for** $j = 1$ to $N$ **do**
9:         $F_{\text{emp}}(x_{\text{proj}}^{j,i}) = \frac{j}{N}$         ▷ Empirical Cumulative Distribution Function (ECDF)
10:         $F_{\text{ref}}(x_{\text{proj}}^{j,i}) = \int_{-\infty}^{x_{\text{proj}}^{j,i}} f(x)\,dx$         ▷ CDF of the Reference Gaussian Distribution
11:         $D^{j,i} = |F_{\text{emp}}(x_{\text{proj}}^{j,i}) - F_{\text{ref}}(x_{\text{proj}}^{j,i})|$         ▷ K-S Statistic
12:     **end for**
13:     $D^i = \max_j D^{j,i}$
14:     **if** $D^i$ is greater than the critical value at significance $\alpha$ for dimension $i$ **then**
15:         Reject $H_0$ for dimension $i$
16:     **else**
17:         Do not reject $H_0$ for dimension $i$
18:     **end if**
19: **end for**

---

To determine if GMM is appropriate to model the scRNA-seq data, we apply Kolmogorov-Smirnov (K-S) Test to all principal components as detailed in Algorithm 2. If the K-S test suggests that the data is not drawn from a single Gaussian, this can be an initial hint (though not definitive proof) that a GMM might be suitable. Although this approach may not capture the true multivariate nature of the data or account for inter-feature dependencies, it is useful in identifying the presence of multiple modes or clusters in the data.

## A.6 MAP Parameter Estimation

---

**Algorithm 3** MAP Parameter Estimation for the GMM via the EM Algorithm

---

1: **Input:** $\{k, \boldsymbol{x}\}$. $k$ denotes the number of Gaussian components. $\boldsymbol{x} \in \mathbb{R}^{N \times m}$ denotes $N$ samples, each with $m$ dimensions after PCA.
2: **Output:** $\{\boldsymbol{\pi}_{1:k}, \boldsymbol{\mu}_{1:k}, \boldsymbol{\Sigma}_{1:k}\}$ MAP parameter estimates, where $\boldsymbol{\pi}_k$ is the prior probability, $\boldsymbol{\mu}_k \in \mathbb{R}^m$ is the mean vector, and $\boldsymbol{\Sigma}_k \in \mathbb{R}^{m \times m}$ is the covariance matrix of component $k$.
3: Initialize $\boldsymbol{\pi}_i, \boldsymbol{\mu}_i, \boldsymbol{\Sigma}_i$ for $i = 1 : k$.
4: **repeat**
5:     **for** $i = 1 : k$ **do**                      ▷ Iterate through each Gaussian component in the mixture.
6:         **for** $j = 1 : N$ **do**             ▷ Calculate log-likelihood & posterior for each sample.
7:             $\log P(\boldsymbol{x}_j \mid \boldsymbol{\pi}_i) = -\frac{m}{2}\log(2\pi) - \frac{1}{2}\log|\boldsymbol{\Sigma}_i| - \frac{1}{2}(\boldsymbol{x}_j - \boldsymbol{\mu}_i)^T \boldsymbol{\Sigma}_i^{-1}(\boldsymbol{x}_j - \boldsymbol{\mu}_i)$
8:             $a_{i,j} = \log P(\boldsymbol{x}_j \mid \boldsymbol{\pi}_i) + \log P(\boldsymbol{\pi}_i)$
9:             $\log P(\boldsymbol{x}_j) = \log \sum_{l=1}^{k} \exp(a_{l,j} - \max_l a_{l,j}) + \max_l a_{l,j}$    ▷ Log-Sum-Exp trick
10:             $P(\boldsymbol{\pi}_i \mid \boldsymbol{x}_j) = \exp(a_{i,j} - \log P(\boldsymbol{x}_j))$                  ▷ E-Step
11:         **end for**
12:         $\boldsymbol{\mu}_i = \frac{\sum_{j=1}^{N} P(\boldsymbol{\pi}_i | \boldsymbol{x}_j) \boldsymbol{x}_j}{\sum_{j=1}^{N} P(\boldsymbol{\pi}_i | \boldsymbol{x}_j)}$
13:         $\boldsymbol{\Sigma}_i = \frac{\sum_{j=1}^{N} P(\boldsymbol{\pi}_i | \boldsymbol{x}_j)(\boldsymbol{x}_j - \boldsymbol{\mu}_i)(\boldsymbol{x}_j - \boldsymbol{\mu}_i)^T}{\sum_{j=1}^{N} P(\boldsymbol{\pi}_i | \boldsymbol{x}_j)}$             ▷ M-Step
14:         $\boldsymbol{\pi}_i = \frac{1}{N} \sum_{j=1}^{N} P(\boldsymbol{\pi}_i \mid \boldsymbol{x}_j)$            ▷ Update the priors for each component
15:         $\alpha = 0.01 \times \text{mean}(\text{diag}(\boldsymbol{\Sigma}_i))$
16:         $\boldsymbol{\Sigma}_i = \boldsymbol{\Sigma}_i + \alpha \mathbb{I}$                          ▷ Tikhonov regularization
17:     **end for**
18: **until** Convergence or maximum number of EM iterations

---

We introduce a bias to stabilize the covariance matrix by adding a scaled identity matrix:

$$\tilde{\Sigma} = \Sigma + \alpha I \tag{6}$$

where $\alpha$ is a small positive regularization parameter and $I$ is the identity matrix of the same dimension as $\Sigma$. The choice $\alpha = 0.01 \times \text{mean}(\text{diag}(\Sigma_i))$ is a heuristic wherein the covariance matrix is regularized by adding 1% of its average variance to its diagonal. This represents an arbitrary yet small perturbation. Benefits of covariance regularization include:

- **Invertibility:** When the number of dimensions is close to or exceeds the number of data points, $\Sigma$ might be singular. Regularization ensures its invertibility.

- **Stability:** For ill-conditioned matrices, their inversion can be highly sensitive to slight changes in the data. Regularization stabilizes the inversion process. The condition number is a commonly used measure to gauge the stability of a matrix, especially when it comes to measuring how a matrix will amplify errors in problems involving matrix inversion. The condition number of a matrix $A$ in terms of its norm is defined as:

$$\kappa(A) = \|A\| \cdot \|A^{-1}\|$$

  For the 2-norm (or Euclidean norm), it can be expressed in terms of the singular values $\sigma_{\max}$ and $\sigma_{\min}$ of $A$ as:

$$\kappa_2(A) = \frac{\sigma_{\max}}{\sigma_{\min}}$$

  Given a linear system $Ax = b$:
  If $\kappa(A) \approx 1$, then $A$ is well-conditioned. For small relative changes in $b$ or $\delta A$, the corresponding changes in the solution $x$ are also small.
  If $\kappa(A) \gg 1$, then $A$ is ill-conditioned. Minor relative perturbations in $b$ or $\delta A$ can lead to significant variations in the solution $x$.

- **Eigenvalue Shrinkage:** The regularization effectively increases each eigenvalue of $\Sigma$ by $\alpha$, beneficial when there is a need to dampen the influence of high variance variables.

### A.7 Experimental Results

Table 3: Quantitative Analysis of Gene Perturbation Responses in CRISPRi Experiments. This table presents the effectiveness of sc-OTGM in identifying known differentially expressed genes (DEGs) following targeted gene knockdowns. Statistical validation is provided through $p$-Values obtained from Fisher's exact test, confirming a significant correlation between predicted and known DEGs for all targeted gene knockdown experiments, well below the significance threshold of 0.05. The columns labeled 'Accuracy (%)' and 'F1-score' evaluate sc-OTGM's performance to accurately predict the direction of expression changes—either upregulation or downregulation—in DEGs. The mean accuracy of $75.2\% \pm 15.4\%$ and mean F1-score of $0.79 \pm 0.14$ show the model's performance in predicting gene expression dynamics.

| Gene | DEG Enrichment | Perturbation Response Prediction | |
|---|---|---|---|
| | $p$-Value | Acc. (%) | F1-score |
| TUBB4A | $7.37 \times 10^{-6}$ | 100.0 | 1.00 |
| ATP1A3 | $3.12 \times 10^{-20}$ | 78.7 | 0.85 |
| KIFAP3 | $6.39 \times 10^{-10}$ | 87.5 | 0.67 |
| MAPT | $9.01 \times 10^{-4}$ | 100.0 | 1.00 |
| CASP3 | $6.93 \times 10^{-6}$ | 100.0 | 1.00 |
| APEX1 | $2.61 \times 10^{-16}$ | 100.0 | 1.00 |
| COX10 | $1.43 \times 10^{-16}$ | 83.7 | 0.87 |
| NDUFS8 | $2.29 \times 10^{-36}$ | 82.7 | 0.83 |
| ZNF292 | $4.91 \times 10^{-16}$ | 65.4 | 0.72 |
| GSTA4 | $5.50 \times 10^{-21}$ | 89.5 | 0.92 |
| STX1B | $4.95 \times 10^{-38}$ | 72.1 | 0.76 |
| OPTN | $1.51 \times 10^{-6}$ | 100.0 | 1.00 |
| SOD1 | $1.16 \times 10^{-7}$ | 88.9 | 0.90 |
| NDUFV1 | $4.33 \times 10^{-30}$ | 73.9 | 0.79 |
| CALB1 | $1.38 \times 10^{-4}$ | 40.0 | 0.46 |
| EEF2 | $5.08 \times 10^{-29}$ | 92.2 | 0.92 |
| BIN1 | $6.39 \times 10^{-8}$ | 88.9 | 0.89 |
| SCFD1 | $1.59 \times 10^{-42}$ | 56.4 | 0.66 |
| PON2 | $8.41 \times 10^{-55}$ | 53.6 | 0.60 |
| BAX | $1.27 \times 10^{-30}$ | 78.3 | 0.83 |
| SCAPER | $1.48 \times 10^{-31}$ | 87.2 | 0.89 |
| CYB561 | $5.66 \times 10^{-33}$ | 60.4 | 0.66 |
| AKAP9 | $3.69 \times 10^{-14}$ | 100.0 | 1.00 |
| VPS35 | $9.36 \times 10^{-14}$ | 80.0 | 0.85 |
| PRNP | $3.95 \times 10^{-53}$ | 59.4 | 0.72 |
| AP2A2 | $2.48 \times 10^{-57}$ | 75.4 | 0.82 |
| SOD2 | $2.01 \times 10^{-7}$ | 90.5 | 0.91 |
| BECN1 | $6.22 \times 10^{-4}$ | 73.1 | 0.75 |
| SNCB | $6.79 \times 10^{-41}$ | 87.7 | 0.89 |
| CDH11 | $1.77 \times 10^{-14}$ | 66.7 | 0.70 |
| ELOVL5 | $2.28 \times 10^{-14}$ | 92.3 | 0.93 |
| NTRK2 | $4.50 \times 10^{-29}$ | 58.7 | 0.61 |
| DAP | $1.27 \times 10^{-45}$ | 81.7 | 0.86 |
| EIF4G1 | $3.00 \times 10^{-24}$ | 76.4 | 0.84 |
| TRPM7 | $3.66 \times 10^{-14}$ | 66.7 | 0.80 |
| COASY | $1.51 \times 10^{-6}$ | 83.3 | 0.84 |
| TRAP1 | $2.11 \times 10^{-35}$ | 80.0 | 0.89 |
| CYP46A1 | $4.56 \times 10^{-4}$ | 50.0 | 0.33 |
| PARP1 | $5.51 \times 10^{-24}$ | 57.7 | 0.67 |
| FOXRED1 | $1.49 \times 10^{-25}$ | 75.8 | 0.78 |
| AFG3L2 | $1.01 \times 10^{-15}$ | 75.0 | 0.81 |
| RAB7A | $1.21 \times 10^{-12}$ | 83.3 | 0.84 |
| PPP2R2B | $3.10 \times 10^{-35}$ | 63.9 | 0.75 |
| RGS2 | $5.03 \times 10^{-30}$ | 63.0 | 0.68 |
| AMFR | $4.06 \times 10^{-12}$ | 62.5 | 0.74 |
| MRPL10 | $1.11 \times 10^{-22}$ | 43.6 | 0.57 |

Table 3 – *Continued from previous page*

| Gene | DEG Enrichment | Perturbation Response Prediction | |
|---|---|---|---|
| | $p$-Value | Acc. (%) | F1-score |
| ANO10 | $1.38 \times 10^{-4}$ | 40.0 | 0.46 |
| DMXL1 | $6.72 \times 10^{-33}$ | 66.7 | 0.76 |
| HYOU1 | $7.43 \times 10^{-35}$ | 55.3 | 0.65 |
| HTT | $8.81 \times 10^{-24}$ | 61.1 | 0.66 |
| ECHS1 | $5.44 \times 10^{-9}$ | 71.4 | 0.79 |
| CYCS | $7.80 \times 10^{-3}$ | 77.8 | 0.80 |
| CEP63 | $2.80 \times 10^{-14}$ | 75.0 | 0.77 |
| FARP1 | $1.35 \times 10^{-22}$ | 83.3 | 0.88 |
| FRMD4A | $8.53 \times 10^{-31}$ | 83.1 | 0.84 |
| RPL6 | $2.74 \times 10^{-36}$ | 67.7 | 0.77 |
| PFN1 | $3.91 \times 10^{-16}$ | 80.0 | 0.85 |

## A.8 Implementation of Hit-and-Run Markov Chain Sampler

We implemented this code to simulate the performance of Hit-and-Run sampling in estimating the cross-covariance matrix between two distributions. Two sets of parameters (means and covariances) for two different multivariate normal distributions are used to generate two sets of samples: `base_samples` and `noise`. `base_samples` are generated using the first set of parameters (`mean1`, `cov1`) by sampling from the corresponding multivariate normal distribution. Similarly, `noise` is generated using the second set of parameters (`mean2`, `cov2`). To create a correlation between these two sets of samples, the function uses a `correlation_factor`. This factor determines how much of the final synthetic data (`correlated_samples`) will be influenced by the `base_samples` versus the `noise`. This introduces a controlled amount of dependence between the two sets of samples, simulating correlated data. By controlling the `correlation_factor`, the code can simulate different degrees of correlation between the two sets of data. This synthetic data is then used in the code to evaluate the performance of the Hit-and-Run sampler in estimating the cross-covariance of correlated distributions. It evaluates the accuracy of the estimation by comparing it to a ground truth and measures performance using Root Mean Square Error (RMSE) and Frobenius norm.

```python
import numpy as np
from scipy.stats import multivariate_normal, norm
import matplotlib.pyplot as plt

def generate_random_params(dimension: int) -> tuple:
    """
    Generate random means and covariances for a given dimension.

    Args:
    dimension (int): The dimensionality of the mean and covariance.

    Returns:
    tuple: A tuple containing the mean vector and covariance matrix.
    """
    mean = np.random.randn(dimension)
    random_matrix = np.random.randn(dimension, dimension)
    # Symmetric, positive-definite matrix
    cov = np.dot(random_matrix, random_matrix.T)
    # Small positive value to the diagonal for numerical stability
    cov += np.eye(dimension) * 1e-6
    return mean, cov

def generate_non_independent_data(
    mean1: np.ndarray,
    cov1: np.ndarray,
    mean2: np.ndarray,
    cov2: np.ndarray,
    num_samples: int,
    correlation_factor: float,
) -> tuple:
    base_samples = np.random.multivariate_normal(mean1, cov1, num_samples)
    noise = np.random.multivariate_normal(mean2, cov2, num_samples)
    correlated_samples = base_samples * correlation_factor + noise * (
        1 - correlation_factor
    )
    return base_samples, noise, correlated_samples

def get_non_zero_density_bounds(
    cov: np.ndarray,
    point: np.ndarray,
    direction: np.ndarray,
    confidence: float = 0.95,
) -> tuple:
    """
    Calculate the bounds for non-zero density in a specified direction.

    Args:
        cov (np.ndarray): Covariance matrix of the distribution.
        point (np.ndarray): Current point in the distribution.
        direction (np.ndarray): Direction vector for calculating bounds.
        confidence (float, optional): Confidence level for the bounds.
            Default is 0.95.

    Returns:
    tuple: A tuple of lower and upper bounds.
    """
    # Normalize the direction vector
    direction_normalized = direction / np.linalg.norm(direction)

    # Project the point onto the direction
    point_projection = np.dot(point, direction_normalized)
```

```python
        # Calculate the variance of the projection
        variance_projection = np.dot(
            direction_normalized.T, np.dot(cov, direction_normalized)
        )
        # Calculate the standard deviation of the projection
        std_deviation = np.sqrt(variance_projection)
        # Find the z-scores for the specified confidence interval
        z_score = norm.ppf((1 + confidence) / 2)

        # Calculate the bounds
        lower_bound = point_projection - z_score * std_deviation
        upper_bound = point_projection + z_score * std_deviation
    return lower_bound, upper_bound

def estimate_cross_covariance(
    mean1: np.ndarray,
    cov1: np.ndarray,
    mean2: np.ndarray,
    cov2: np.ndarray,
    ground_truth: np.ndarray,
    num_iterations: int = 1000,
) -> tuple:
    """
    Estimate the cross-covariance matrix using Hit-and-Run Markov chain
    sampling and calculate the RMSE and Frobenius norm at each iteration.

    Args:
        mean1, mean2 (np.ndarray): Mean vectors of the first and
            second distributions.
        cov1, cov2 (np.ndarray): Covariance matrices of the first and
            second distributions.
        ground_truth (np.ndarray): Ground truth cross-covariance matrix.
        num_iterations (int, optional): Number of iter. for sampling.
            Default is 1000.

    Returns:
        tuple: Estimated cross-covariance matrix, list of RMSE values,
            and list of Frobenius norms.
    """
    rmse_values, frobenius_norms = [], []

    # Random initialization of cross-covariance matrix
    dimension = mean1.shape[0]
    random_matrix = np.random.randn(dimension, dimension)
    cross_covariance = (
        np.dot(random_matrix, random_matrix.T) + np.eye(dimension) * 1e-6
    )

    # Initialize starting points
    x_current = np.random.multivariate_normal(
        mean1, np.diag(np.ones(dimension))
    )
    y_current = np.random.multivariate_normal(
        mean2, np.diag(np.ones(dimension))
    )

    for i in range(1, num_iterations + 1):
        # Hit-and-Run sampling for X
        direction_x = np.random.randn(len(mean1))
        lb_x, ub_x = get_non_zero_density_bounds(
            cov1, x_current, direction_x
        )
        x_current = np.random.uniform(lb_x, ub_x)
```

```python
        # Hit-and-Run sampling for Y
        direction_y = np.random.randn(len(mean2))
        lb_y, ub_y = get_non_zero_density_bounds(
            cov2, y_current, direction_y
        )
        y_current = np.random.uniform(lb_y, ub_y)

        # Sequential update on cross-covariance matrix
        deviation_x = x_current - mean1
        deviation_y = y_current - mean2
        cross_covariance = (i / (i + 1)) * cross_covariance + (
            i / (i + 1) ** 2
        ) * np.outer(deviation_x, deviation_y)

        # Update rmse values and frobenius_norms
        rmse = np.sqrt(np.mean((cross_covariance - ground_truth) ** 2))
        rmse_values.append(rmse)
        fro_norm = np.linalg.norm(cross_covariance - ground_truth, "fro")
        normalized_fro_norm = fro_norm / np.sqrt((dimension**2))
        frobenius_norms.append(normalized_fro_norm)

    return cross_covariance, rmse_values, frobenius_norms

def main(
    dimension: int,
    num_iterations: int = 1000,
    correlation_factor: float = 0,
) -> list:
    """
    Process and estimate cross-covariance for a given dimension with an
    imposed correlation factor.

    Args:
        dimension (int): The dimensionality of the distribution.
        num_iterations (int, optional): Number of iter. for sampling.
            Default is 1000.
        correlation_factor (float, optional): Factor to impose
            correlation between the two distributions. Default is 0.

    Returns:
        list: List of RMSE values over iterations and list of Frobenius
            norms for iterations.
    """
    mean1, cov1 = generate_random_params(dimension)
    mean2, cov2 = generate_random_params(dimension)

    # Generate correlated samples
    _, _, correlated_samples = generate_non_independent_data(
        mean1, cov1, mean2, cov2, num_iterations, correlation_factor
    )

    # Calculate the ground truth cross-covariance matrix
    ground_truth = np.cov(correlated_samples, rowvar=False)

    _, rmse, frobenius_norms = estimate_cross_covariance(
        mean1, cov1, mean2, cov2, ground_truth, num_iterations
    )
    return rmse, frobenius_norms

if __name__ == "__main__":
    dimensions = [100]
    num_iterations = 1000
```

```python
correlation_factors = [0.0, 0.1, 0.2, 0.5]

# Dictionary to store results for each correlation factor
all_results = {}

# Compute results for each correlation factor and dimension
for factor in correlation_factors:
    all_results[factor] = {
        dim: main(dim, num_iterations, factor) for dim in dimensions
    }

plt.figure(figsize=(16, 5))
lines = []   # To store line objects for the legend
labels = []  # To store label strings for the legend

for i, factor in enumerate(correlation_factors):
    plt.subplot(2, 4, i + 1)

    # Retrieve results for this correlation factor
    results = all_results[factor]

    # Plotting RMSE values
    for dim, (rmse, _) in results.items():
        (line,) = plt.plot(rmse, label=f"d={dim}", linewidth=1)
        if i == 0:  # Only add to legend for the first subplot
            lines.append(line)
            labels.append(f"d={dim}")

    plt.title(f"$\\rho$: {factor}", fontsize=12)
    plt.xlabel("# of iterations", fontsize=10)
    plt.ylabel("RMSE", fontsize=10)
    plt.xlim(0, num_iterations)
    plt.grid(True)

for i, factor in enumerate(correlation_factors):
    plt.subplot(2, 4, i + 5)

    # Retrieve results for this correlation factor
    results = all_results[factor]

    # Plotting Frobenius norms
    for dim, (_, fro_norm) in results.items():
        plt.plot(fro_norm, label=f"d={dim}", linewidth=1)

    plt.title(f"$\\rho$: {factor}", fontsize=12)
    plt.xlabel("# of iterations", fontsize=10)
    plt.ylabel("Frobenius norm", fontsize=10)
    plt.xlim(0, num_iterations)
    plt.grid(True)

plt.figlegend(
    lines,
    labels,
    loc="upper center",
    ncol=len(dimensions),
    fontsize=10,
)
plt.tight_layout(rect=[0, 0, 1, 0.95])
plt.show()
```

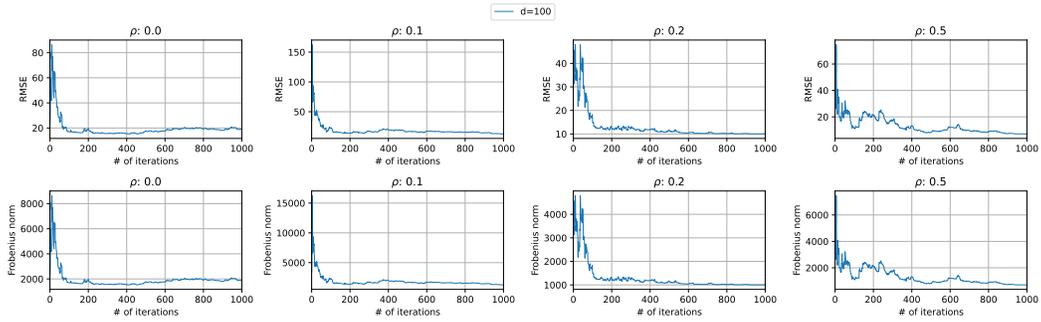

(a) $d = 100$: Convergence trends at high-dimensional setting.

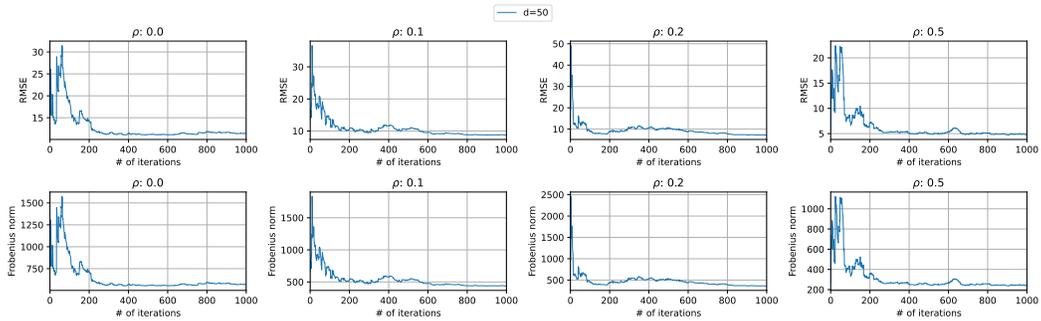

(b) $d = 50$: Mid-dimensional performance.

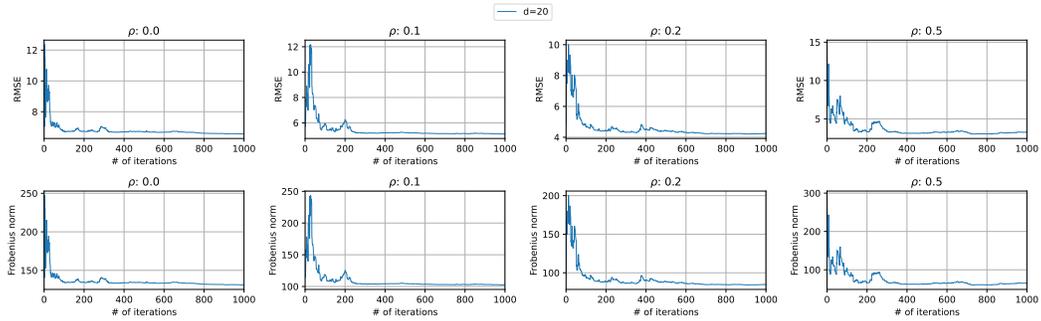

(c) $d = 20$: Comparative analysis in a moderate/low dimensional space.

Figure 4: Performance Analysis of the Hit-and-Run Markov Chain Sampler on Synthetic Data. The plots demonstrate the convergence behavior of the sampler in terms of RMSE and Frobenius norm over 1000 iterations, across varying levels of correlation factor $\rho$ ranging from 0.0 to 0.5. These results show the efficiency and accuracy of the sampler for different dimensions $d$ of the synthetic data.